\documentclass[11pt,tightenlines,onecolumn, aps, prd, nofootinbib, superscriptaddress, showkeys, showpacs]{revtex4-1}

\usepackage{latexsym}
\usepackage{amssymb}
\usepackage[dvips]{graphicx}
\usepackage{color}
\usepackage{graphicx}
\usepackage{float}
\usepackage[font=small]{caption}

\usepackage{amsmath, amsfonts, amssymb, mathrsfs}
\usepackage{amsthm}
\usepackage{tensor}
\usepackage{multirow}
\usepackage{bbm}
\usepackage[geometry]{ifsym}

\usepackage{tensor}

\newcommand{\minkowski}{\tensor{\eta}{_\mu_\nu}}

\newcommand{\metric}{\tensor{g}{_\mu_\nu}}

\newcommand{\hmn}{\tensor{h}{_\mu_\nu}}

\newcommand{\Pg}{\,{\Psi}_g\,}
\newcommand{\Og}{{\Omega}_g\,}
\newcommand{\cPg}{\,\overline{\Psi}_g\,}
\newcommand{\cOg}{\overline{\Omega}_g\,}

\newcommand{\Dmu}{\,\Delta_\mu\,}
\newcommand{\Dnu}{\,\Delta_\nu\,}
\newcommand{\Pmu}{\,\Pi_\mu\,}
\newcommand{\Pnu}{\,\Pi_\nu\,}

\newcommand{\gc}{\tensor{\gamma}{_5}}

\newcommand{\gmct}{\tensor{\gamma}{^\mu}}
\newcommand{\gnct}{\tensor{\gamma}{^\nu}}

\newcommand{\Jcov}{\tensor{J}{_\mu}}
\newcommand{\Jct}{\tensor{J}{^\mu}}

\begin{document}

\title{Gravitational waves in the Spinor Theory of Gravity}
\author{M. Novello}

\affiliation{ Centro de Estudos Avan\c{c}ados de Cosmologia (CEAC/CBPF) \\
 Rua Dr. Xavier Sigaud, 150, CEP 22290-180, Rio de Janeiro, Brazil. }

\author{A.~E.~S. Hartmann}
\affiliation{Dipartamento di Scienza e Alta Tecnologia, Universit\`{a} degli Studi dell'{}Insubria,  \\
	Via Valleggio 11, 22100 Como, and INFN, Sez di Milano, Italy.}
\date{\today}

\vspace{0.50cm}

\begin{abstract}
We analyze the gravitational waves within the Spinor Theory of Gravity and compare it with the General Relativity proposal. In the case of STG a gravitational wave may occur if the effective gravitational metric induced by the spinorial field is Ricci flat.
\end{abstract}


 \maketitle

\vspace{0.50cm}

Recently we have been discussing the possibility of to represent gravitation by two spinor fields $\Pg$ and $\Omega_g$ interacting through weak process in the Minkowski spacetime \cite{novhart20}. For the action of the gravitational neutrinos (g-neutrinos for short) is stated as \cite{note1}
\begin{align}
S = S_D \,+\, S_g\,,
\end{align}
with
\begin{align}
S_D = \int \sqrt{-\eta}\,\left( \frac{i}{2}\cPg\gmct\nabla_\mu\Pg -  \frac{i}{2} (\overline{\nabla}_\mu\cPg)\gmct\Pg +  \frac{i}{2}\cOg\gmct\nabla_\mu\Og - \frac{i}{2}(\overline{\nabla}_\mu\cOg)\gmct\Og   \right)d^4x
\end{align}
and
\begin{align}
S_g = \int \sqrt{-\eta}\,\minkowski\bigl[ \cPg\gmct(1-\gc)\Og \bigr] \bigl[ \cOg\gnct(1-\gc)\Pg \bigr]\,d^4x\,.  \label{g-action} 
\end{align}
The covariant derivative $ \nabla_{\mu}$ is defined as
$$ \nabla_{\mu}   \Psi = \partial_{\mu}  \Psi - \Gamma_{\mu} \, \Psi,$$
with the general internal connection $\Gamma_{\mu} = \Gamma^{FI}_{\mu} + U_\mu$.

By gravitation we call the modification of the flat background to a curved riemannian geometry $\metric$, represented in the covariant form as
\begin{align}
\metric = \minkowski - \kappa\,\hmn\,,\label{metric}
\end{align}
where $\kappa$ is the Einstein's constant and the $\hmn$ are constructed with two null-vectors $\Delta_\mu$ and $\Pi_\mu$  defined in terms of the V--A weak current and carry dimension $ L^{- 1}$, namely
\begin{align}
\hmn = (\Dmu +\,\Pmu)(\Dnu +\, \Pnu) \label{hmn}
\end{align}
and \cite{note2}
\begin{align}
\Dmu &\equiv \left(\frac{g_w}{J^2}\right)^{1/4} \, \left( J_\mu - I_\mu \right)\,, \\[1.5ex]
\Pmu &\equiv \left(\frac{g_w}{J^2}\right)^{1/4} \, \left( j_\mu - i_\mu \right)\,.
\end{align}
By effective \cite{effective} we mean $\metric$ has not a dynamics by its own as in GR, but instead inherits its dynamical evolution from the g-neutrino fields. The main motivation concerning the relation between gravity and Fermi process is expressed by the dimensionality of its constants $\kappa$ and $g_w$, respectivelly (see the discussion presented in \cite{novhart20} and the references therein).

At this point, mention should be made of the fact that the inverse metric is as exact as the covariant expression (\ref{metric}), due to the null vectors $\Dmu$ and $\Pmu$, which allows us to write the following closure relation
\begin{align}
\tensor{h}{_\mu_\alpha}\,\tensor{h}{^\alpha_\nu} = 2\kappa\,\Delta_\alpha\,\Pi^\alpha\, \hmn\,. \label{cr}
\end{align}
Besides there is a scalar field $ H $ introduced via the internal connection that participate in the dynamics of these g-neutrinos. A more detailed presentation of this framework is made in \cite{novhart20}. The present communication intends to discuss how gravitational waves can be described in a restricted case of STG -- that is when only one g-neutrino modifies the Minkowskian background. In this case, the expression for $\hmn$, equations (\ref{hmn}) and (\ref{cr}), becomes simply
\begin{align}
h_{\mu\nu} &= \,\Delta_{\mu} \, \Delta_{\nu},\label{hmn'}\\
\tensor{h}{_\mu_\alpha}\,\tensor{h}{^\alpha_\nu} &= 0. \label{cr'}
\end{align}

In general, a classical physical wave is defined through the discontinuity of the equation of motion of the field and/or its derivatives. The most useful formulation of propagation of waves was done by J. Hadamard \cite{hadamard}. 
Once in STG $\metric$ has not a dynamics by its own, this approach can not be followed. Instead a exact solution for the \textit{c} field $\Psi_g$ allows us to implement it in STG.


Let us briefly comment on the definition of gravitational waves \cite{zakharov} in the context of General Relativity (GR). For the simplest description is constructed considering the linear approximation of the metric under the form \cite{yvonne}
\begin{equation}
g_{\mu\nu} \approx \eta_{\mu\nu} + \epsilon \, \phi_{\mu\nu}
\label{1jlho7}
\end{equation}
The vacuum wave equations are obtained from Einstein's equations by assuming that the weak field is traceless and divergence-free,
\begin{align}
\eta^{\mu\nu}\,\phi_{\mu\nu} &=0,\\
\tensor{\phi}{_\mu^\nu_{,\nu}} &= 0. \label{1julho6}
\end{align}
Then the Ricci-flat spacetime $R_{\mu\nu}=0$ is achieved if
\begin{equation}
\Box_{\eta} \, \phi_{\mu\nu} = 0.
\label{1julho5}
\end{equation}
The subscript $\eta$ denotes the D'Alembertian in the Minkowski background. A solution of (\ref{1julho5}) is called gravitational wave in GR and propagates with the velocity of light. 

The analysis of $\phi_{\mu\nu}$ in the case of arbitrary (not weak) field can be most directly presented if one selects to work in the harmonic coordinate system
\begin{equation}
 \Gamma^{\alpha}_{\mu\nu} \, g^{\mu\nu} = 0,
 \label{10julho1}
 \end{equation}
in which the dynamics of GR becomes  \cite{yvonne} 
\begin{equation}
g^{00} \, g_{i j ,00} + \Sigma_{ij} = 0
\label{9julho1}
\end{equation}
where  $ \Sigma_{ij} $ does not contain second time derivatives and thus are determined by Cauchy data. Then it is immediate to show that if there is a discontinuity $ \chi$  it is characterized by the eikonal
\begin{equation}
g^{\mu\nu} \, \chi_{, \mu} \, \chi_{, \nu} = 0.
\label{9julho2}
\end{equation}

Let us turn now to the case of STG.




The dynamics of the g-neutrino $ \Psi_g$ (an analogous equation is valid for the other g-neutrino is given by the Dirac equation
\begin{equation}
 i \, \gamma^{\mu} \, \nabla_{\mu} \Psi_g = 0.
\label{Dirac}
\end{equation}

Note that the g-neutrino lives in the flat Minkowski space-time and if we choose an Euclidean coordinate system
$$ ds^2 = dt^2 - dx^2 - dy^2 - dz^2, $$
the gamma matrices $\gamma_\mu$ are constant and the covariant derivative of the spinors is controlled by the scalar field $ H$, namely \cite{novhart20}
$$ \Gamma_{\mu} =  \frac{1}{4} \, \gamma_{\mu} \, \gamma^{\alpha} \, H_{, \alpha}.$$
A comma means simple derivative and $ H$ obeys the Klein-Gordon equation
$$ \Box H \equiv \eta^{\mu\nu} \, \partial_{\mu} \, \partial_{\nu} \,H = 0.$$

It then follows that it is possible to substitute in these formulas $ \eta_{\mu\nu}$ by $ g_{\mu\nu}.$ Furthermore, we note that due to the limitation to just one single spinor field, combining with the property that $ \Delta_{\mu}$ is a null vector imply that the determinant of $ g_{\mu\nu} $ is equal to the determinant of $ \eta_{\mu\nu}.$ In \cite{novhart20} it is shown that this limited framework contains two relevant metrics -- the spherically symmetric and static field of a massive object and the dynamical isotropic universe.

\vspace{.5cm}
A solution of (\ref{Dirac}) is provided by setting
\begin{equation}
\Psi_g = \exp H \, \Psi^{o},
\label{1julho1}
\end{equation}
with $ \Psi^{o}$ is a constant spinor. The corresponding null-vector of the Fermi (weak) current is
$$ \Delta_{\mu} =  \exp H  \, n_{\mu} $$
where $ n_{\mu}$ is a constant null vector in Minkowski space-time,  whose respective derivative yields
\begin{equation}
 \Delta_{\mu , \alpha} = H_{, \alpha} \, \Delta_{\mu}
\label{11julho1}
\end{equation}
It is immediate to see that in the effective metric $ g_{\mu\nu} = \eta_{\mu\nu}- \kappa \, \Delta_{\mu} \, \Delta_{\nu} $ the covariant derivative of the null vector  $\Delta_{\mu} $ takes the same form
\begin{align}
\Delta_{\mu \, ; \lambda} =  \Delta_{\mu \, , \lambda}.
\end{align}

Let us consider the case in which the vector $ H_{\mu} $ is orthogonal to $\Delta_{\mu}.$ A simple possibility is given by setting
\begin{equation}
 H = e^{ \, n_{\mu} x^{\mu}}.
 \label{11julho5}
 \end{equation}
Then, a direct calculation show that this implies the Fock harmonic condition (\ref{10julho1}). Consequently, $ h_{\mu\nu}$ satisfies the exact equations
\begin{equation}
\Box_{g} \, h_{\mu\nu} = 0.
\label{1julho8}
\end{equation}
\begin{equation}
h^{\mu\nu}{}_{;  \, \nu} = 0.
\label{1julho9}
\end{equation}
\begin{equation}
 h_{\mu\nu} \, g^{\mu\nu} =0.
 \label{1julho10}
 \end{equation}
The symbol $ g $ means that the (covariant) derivative must be taken in the geometry driven by the effective metric
$$g_{\mu\nu} = \eta_{\mu\nu} - \kappa \, \exp(2H) \, n_{\mu} \, n_{\nu}.$$
Note that although similar in the form of the linear case of GR, the above equations for STG are not approximations but exact. 
Moreover $ n_{\mu}$ is a null vector in Minkowski and also in the effective gravitational metric. Such properties allows us to propose that equations (\ref{1julho8}, \ref{1julho9}, \ref{1julho10}) should be taken as the definition of gravitational waves in STG.
 
\vspace{.5cm}
Finally, let us point out that the properties satisfied by the scalar field $ H $ are similar to that ones required by W. Kundt \cite{kundt} in his definition of gravitational waves in GR. Indeed, for Kundt the metric of a space-time $ V_{4}$ will describe a field of plane waves if the given $ V_{4}$ admits an isotropic vector field $ P^{\alpha}$ satisfying the conditions
\begin{itemize}
\item{$P_{[\alpha ; \beta]} = 0;$}
\item{$P_{(\alpha ; \beta)} \, P^{\alpha ; \beta} = 0;$}
\item{$P^{\alpha}{}_{; \alpha} = 0.$}
\end{itemize}
It is immediate to verify that identifying $ P^{\alpha}$ to the gradient of the scalar field $ H_{, \alpha}$ is enough to fulfil the three Kundt requirements.

In order to further characterize the properties of STG with the metric (\ref{metric}) under (\ref{hmn'}), let us evaluate the contracted Riemann curvature $ R_{\mu\nu}$ under the condition (\ref{11julho1}) and the case driven by (\ref{11julho5}). It is immediate to check it from the Christoffel symbols
\begin{align*}
\Gamma^{\alpha}_{\mu\nu} = \frac{\kappa}{2} \, g^{\alpha\beta} \,  \left[ - \, (\Delta_{\mu} \, \Delta_{\beta})_{ , \, \nu} \, - \,  (\Delta_{\nu} \, \Delta_{\beta})_{ , \, \mu} + \,
 (\Delta_{\mu} \, \Delta_{\nu})_{ , \, \beta} \right], 
\end{align*}
where $ g^{\alpha\beta} = \eta^{\alpha\beta} + \kappa\Delta^{\alpha} \, \Delta^{\beta}.$

\vspace{.5cm}

For we can state the following

$$\mathbf{Lemma}$$

\begin{enumerate}
\item{Let $ \Psi_{g}$ be a spinor field in the Minkowski spacetime $ V_{4}$ that modifies the metric through the weak current $ \Delta_{\mu}$ and satisfies the Dirac equation (\ref{Dirac});}
\item{Let $ H$ be the Klein-Gordon field that generalize the internal connection of $ \Psi_{g}$ and whose gradient is orthogonal to  $ \Delta_{\mu}$ ;}
\item{Let the effective metric (\ref{metric}) with (\ref{hmn'}) be the representation of the action of $ \Psi_{g}$ over $ V_{4}$ ;}\\
\\
{Then the effective geometry is Ricci-flat.}
\end{enumerate}

Therefore, the definition of gravitational waves provided by STG with only one g-neutrino requires that the effective metric, as in GR, be Ricci flat. However, unlike GR no gauge issues are implied, the effective geometry from the beginning is exact and, by hypothesis, two dynamical fields are involved -- the g-neutrino $\Psi_g$ and the neutral scalar field $H$.



\begin{thebibliography}{100}

\bibitem{novhart20} M. Novello and A. E. S. Hartmann. From weak interaction to gravity. ArXiv: gr/qc 2006.16810. 

\bibitem{note1} We set $\eta=\det\minkowski$ in an arbritary coordinate system with signature $(+,-,-,-)$. The covariant derivative is $\nabla_\mu = \partial_\mu - \Gamma_\mu$, with $\Gamma_\mu = \Gamma^{FI}_\mu + U_\mu$ being the most general internal connection that satisfies the compatibility with the metric. See the details in \cite{novhart20}.

\bibitem{note2} The g-neutrino currents are $J^\mu = \cPg\,\gmct\,\Pg\,$, $ I^\mu= \cPg\,\gmct\gamma_5\,\Pg\,$; $J^2 = \Jcov\,\Jct$. The same for the $\Og$--currents.

\bibitem{effective} `Effective geometry' is a largely employed concept in analog models of gravity. Its discussion in gravitational physics can be found, for instance, in C. Barcel{\'o}, M. Visser and S. Liberati, Einstein gravity as an emergent phenomenon? \textit{Int. J. of Mod. Phys. D} \textbf{10}, 06 (2001) 799--806.

\bibitem{hadamard} J. Hadamard, \textit{Lectures on Cauchy\rq s Problem}. (Yale University Press, 1923) Dover reprint, 1952.

 \bibitem{zakharov} The old discussion concerning on the criteria for defining gravitational waves in GR is made by V. D. Zakharov, \textit{Gravitational waves in Einstein\rq s theory}. Halsted Press, 1973. A recent update of the subject can be found in S. L. Cacciatori, Gravitational waves, 100 years later. ArXiv: gr/qc 2005.03989. 

\bibitem{yvonne} Y. Choquet-Bruhat. \textit{General Relativity and Einstein\rq s Equations}. Oxford University Press, 2009.



\bibitem{kundt} W. Kundt, The plane-fronted gravitational waves. \textit{Zeitschrift f\"{u}r Physik} \textbf{163} (1961) 77--86.

\end{thebibliography}
 \end{document}